\documentclass[11pt]{article}
\pdfoutput=1 
\usepackage{cite,graphicx,subfigure,xcolor,bm}

\def\Gb{\bar{G}}
\def\Hb{\bar{H}}
\def\Ab{\bar{A}}
\def\Bb{\bar{B}}
\def\Cb{\bar{C}}
\def\I{{\mathcal{I}}}
\def\E{{\mathcal{E}}}
\def\Di{\displaystyle}

\def\eg{{\em e.g.}}

\def\beq{\begin{equation}}
\def\eeq{\end{equation}}
\def\bea{\begin{eqnarray}}
\def\eea{\end{eqnarray}}
\def\D{\mbox{d}}

\def\la{\lambda}

\def\ga{\gamma}
\def\Ga{\Gamma}
\def\Del{\Delta}
\def\inf{\infty}

\def\pa{\partial}

\def\M{\mbox{\scriptsize max}}
\def\m{\mbox{\scriptsize min}}

\def\bold0{\mbox{\boldmath$0$}}

\begin{document}

\begin{center}

{\Large {\bf Accelerated boundary integral analysis of energy eigenvalues
for confined electron states in quantum semiconductor heterostructures}}\\[5mm]

{\bf J. D. Phan$^{\dag}$, A.-V. Phan$^{\ddag}$\footnote{Corresponding author.
\\ $~~~~~\,$ E-mail: \textsf{jphan42@gatech.edu} (J. D. Phan),
\textsf{vphan@southalabama.edu} (A.-V. Phan)}}
\\[3mm]

$^{\dag}\,$Daniel Guggenheim School of Aerospace Engineering \\
Georgia Institute of Technology \\
Atlanta, GA 30332, USA \\[3mm]

$^{\ddag}\,$Department of Mechanical, Aerospace and Biomedical Engineering \\
University of South Alabama \\
Mobile, AL 36688, USA \\[1cm]

\end{center}

\subsection*{Abstract}

This paper presents a novel and efficient approach for the computation of 
energy eigenvalues in quantum semiconductor heterostructures. Accurate 
determination of the electronic states in these heterostructures is crucial 
for understanding their optical and electronic properties, making it a key 
challenge in semiconductor physics. The proposed method is based 
on utilizing series expansions of zero-order Bessel functions to numerically
solve the Schr\"{o}dinger equation using boundary integral method for bound 
electron states in a computationally efficient manner.
To validate the proposed technique, we applied it to address previously 
explored issues by other research groups. The results clearly demonstrate 
the computational efficiency and high precision of our approach. Notably, 
the proposed technique significantly reduces the computational time compared
to the conventional method for searching the energy eigenvalues in quantum 
structures.
\\[-3mm]

\noindent
{\bf Keywords:} Quantum wire; Energy eigenvalue; Boundary integral method;
Wavefunction; Probability density.

\section*{1. Introduction}

In quantum mechanics, the behavior of particles within confined 
structures often leads to interesting and complex physical phenomena.
The study of energy eigenvalues in such systems is not only an interesting
exercise in mathematical 
physics but also holds significant implications for the design and 
understanding of nanoscale devices and quantum structures. This investigation 
utilized boundary element analysis to explore the energy eigenvalues 
associated with the quantum states of various quantum structures,
shedding light on the fundamental principles that govern the quantum behavior 
of particles in complex confined systems.

Although the proposed technique can be applied to quantum heterostructures, 
the primary focus of this research lies in the domain of quantum wires. 
Quantum wire systems have attracted considerable attention in the scientific 
community due to their promising applications in nanoelectronics and quantum 
computing, \eg, \cite{Martin-Palma-17,Lu-23}. Understanding the behavior and 
optimizing the performance of quantum wires relies on the accurate 
determination of their energy eigenvalues.

In this pursuit, researchers commonly employ finite and boundary element 
methods as preferred numerical techniques. For instance, previous studies, 
such as \cite{RM-90,Chen-94,RM-95,Hirayama-01,Mommadi-23}, have 
successfully applied the finite element method for modeling quantum systems.

Boundary element methods (BEMs), on the other hand, offer distinct advantages 
when addressing open-boundary quantum wire systems. BEM is specifically 
tailored to solving the Schr\"{o}dinger equation on the boundary of the 
wire, effectively simplifying the computational domain. Previous research 
endeavors, as demonstrated in references such as \cite{Chen-94,Knipp-94,
Knipp-96,Kosztin-97,Gelbard-01,Gospavic-08,Hohenester-15}, have highlighted 
the use of BEM in this context.

It is essential to note that directly formulating the eigenvalue problem for 
the determination of energy eigenvalues proves to be a complex task. 
Consequently, the approach of scanning the energy variable $E$ within a 
specified interval, in order to locate local minima of the eigenvalue 
determinant, as obtained through boundary element analysis (BEA), has emerged 
as a practical solution. While this search method exhibits a straightforward 
approach, it is not without its limitations. Notably, because the determinant 
is a function of the energy variable $E$, employing this method necessitates 
the complete repetition of the BEA for each value of $E$. This approach, 
though conceptually simple, imposes a substantial computational burden.

When the electron's effective masses remain constant in both the quantum wire 
and barrier regions, the Schr\"{o}dinger equation simplifies to the Helmholtz 
equation. Although the generalized eigenvalue problem was formulated in 
\cite{Gospavic-08}, the technique presented in the paper is limited to 
generating real-valued wavefunctions. This limitation stems from its reliance 
on the real-valued Laplace operator rather than the complex-valued Helmholtz 
operator. In the broader context of quantum systems, it is important to note 
that wavefunctions can take on complex values, enabling them to encompass the 
phase information of the quantum state.

The proposed approach is founded on utilizing series expansions of zero-order 
Bessel 
functions in the fundamental solution to the Helmholtz equation. This novel 
method yields a set of independent fundamental functions that do not rely on 
the wave number of the Helmholtz equation. The resulting coefficient matrix 
in the system of boundary element equations takes the form of a matrix 
polynomial in terms of the parameter $E$. This characteristic enables a 
significantly expedited search for eigenvalues across a specified range of 
$E$ values.

In this work, series expansions of zero-order Bessel functions were employed
in the fundamental solution for the Helmholtz equation. This approach led to 
the development of novel multi-domain boundary integral equations (BIEs) 
comprising a series of boundary integrals that exhibit independence from the 
energy parameter $E$. By implementing these BIEs numerically through boundary 
elements, a system of equations known as the BEA system was formulated. 
Notably, the coefficient matrix of this system takes the form of a matrix
polynomial in the variable $E$.

This innovative technique significantly streamlines the search for eigenvalues 
by scanning the parameter $E$. In contrast to the conventional method, where 
the boundary integrals necessitate re-evaluation for each value of $E$ within 
the specified range, this approach reduces the computational time required.
Several numerical examples were presented to demonstrate the effectiveness 
and accuracy of the proposed technique.

\section*{2. Boundary intergal formulation for energy eigenvalue analysis}

\subsection*{2.1. Boundary integral equations}

For a bound state of a quantum structure, with an interior region denoted 
as $\I$ embedded within a barrier material in the exterior region denoted as 
$\E$, the wavefunctions of the electron are described by the following 
Schr\"{o}dinger equations, \eg, \cite{RM-book}:
\bea
\nabla\cdot\left({1\over m_{\I}^*}\,\nabla\psi^{\I}(Q)\right) +
{2E\over \hbar^2}\,\psi^{\I}(Q) = 0   \label{Schr-I}   \\[2mm]
\nabla\cdot\left({1\over m_{\E}^*}\,\nabla\psi^{\E}(Q)\right) +
{2(E-V_0)\over \hbar^2}\,\psi^{\E}(Q) = 0   \label{Schr-E}
\eea
where $m_{\I}^*$ and $m_{\E}^*$ are the electron's effective masses in 
the interior and exterior regions, respectively, $\hbar$ is the reduced
Planck's constant, $E$ is the energy of the electron and $V_0$ is the
barrier potential.

For constant values of $m_{\I}^*$ and $m_{\E}^*$, the above equations 
reduce to Helmholtz's equations and can
be solved via the following boundary integral equations using their 
respective fundamental solutions:
\bea
c(P)\psi^{\I}(P) &=& \int_{\Ga} \Big[G_{\I}(P,Q)\,\psi^{\I}_{,n}(Q) -
G_{{\I},n}(P,Q)\,\psi^{\I}(Q)\Big]\,\D Q   \label{BIE-I}   \\[2mm]
c(P)\psi^{\E}(P) &=& \int_{\Ga} \Big[G_{\E}(P,Q)\,\psi^{\E}_{,n}(Q) -
G_{{\E},n}(P,Q)\,\psi^{\E}(Q)\Big]\,\D Q   \label{BIE-E}
\eea
where the exterior region is treated as an infinite domain, $P$ and $Q$
represent the source and field points, respectively, $c(P)$ stands for 
the solid angle coefficient \cite{RM-book}, $\Ga$ denotes the interface
that separates the interior
and exterior regions, the subscript $()_{,n}$ signifies the derivative
taken with respect to the unit outward normal $\bm{n}=\bm{n}(Q)$ to $\Ga$, 
\eg, $\psi_{,n}=\Di {\pa\psi\over\pa\bm{n}}$, and $\d Q$ refers to an 
infinitesimal boundary length.

In situations involving two-dimensional (2-D) problems, such as the case of 
a straight quantum wire with a uniform cross-section, the fundamental 
solutions are given by
\bea
G_{\I}(P,Q) &=& {i\over 4}H_0^{(1)}(k_{\I}r) =
-{1\over 4}\Big(Y_0(k_{\I}r) - iJ_0(k_{\I}r)\Big)    \label{GI}   \\[3mm]
G_{\E}(P,Q) &=& {1\over 2\pi}K_0(k_{\E}r) =
-{1\over 4}\Big(Y_0(ik_{\E}r) - iJ_0(ik_{\E}r)\Big)    \label{GE}
\eea

In these equations, $H_0^{(1)}$ is the zero-order Hankel function of the first 
kind, $K_0$ denotes the zero-order modified Bessel function of the second kind,
$J_0$ and $Y_0$ are the zero-order Bessel functions of the first and
second kinds, respectively. Additionally, $i$ represents the imaginary unit, 
and $r$ refers to the distance 
between $P$ and $Q$. The wavenumbers in the interior and exterior 
regions are defined respectively as follows:
\bea
k_{\I}^2 &=& 2m_{\I}^*E/\hbar^2   \\[3mm] 
k_{\E}^2 &=& 2m_{\E}^*(V_0-E)/\hbar^2
\eea
and the interface boundary conditions are
\bea
\psi^{\E} &=& \psi^{\I}   \label{BC1} \\
\psi^{\E}_{,n} &=&-{m_{\E}^*\over m_{\I}^*}\psi^{\I}_{,n} = \mu\psi^{\I}_{,n}
   \label{BC2}
\eea

It should be noted that the boundary integrals in Eqs. (\ref{BIE-I}) and 
(\ref{BIE-E}) require evaluation in the counterclockwise and clockwise 
directions, respectively.

\subsection*{2.2. Numerical implementation}

If the interface boundary $\Ga$ is discretized into $N$ boundary elements 
with a total of $N_n$ boundary nodes, collocating Eqs. (\ref{BIE-I}) and 
(\ref{BIE-E}) at each source point $P=1 \ldots N_n$ respectively yields the 
following equations:
\bea
\sum_{k=1}^{N_n} G^{\I}_{Pk}\psi^{\I}_{,nk} -
\sum_{k=1}^{N_n} H^{\I}_{Pk}\psi^{\I}_k &=& 0   \label{BEM-I} \\
\sum_{k=1}^{N_n} G^{\E}_{Pk}\psi^{\E}_{,nk} -
\sum_{k=1}^{N_n} H^{\E}_{Pk}\psi^{\E}_k &=& 0   \label{BEM-E}
\eea

By using the boundary conditions (\ref{BC1}) and (\ref{BC2}), Eq. 
(\ref{BEM-E}) can be rewritten as
\beq
\mu\sum_{k=1}^{N_n} G^{\E}_{Pk}\psi^{\I}_{,nk} -
\sum_{k=1}^{N_n} H^{\E}_{Pk}\psi^{\I}_k = 0   \label{BEM-E1}
\eeq

The collocation at all $N_n$ source points results in $N_n$ equations of
the form (\ref{BEM-I}) and another $N_n$ equations of the form (\ref{BEM-E1}).
They are two systems of linear equations which can be written in matrix
form as follows:
\bea
\bm{A}(E)\bm{u}_{\I} = \bm{0}   \label{BEM-I2} \\
\bm{B}(E)\bm{u}_{\I} = \bm{0}   \label{BEM-E2}
\eea
or
\beq
\bm{C}(E)\bm{u}_{\I} = \bm{0}   \label{BEM-IE}
\eeq
where $\bm{A}$ and $\bm{B}$ are the coefficient matrices of order 
($N_n\times 2N_n$),
\beq
\bm{C}(E) = \left[ \begin{array}{c}
\bm{A}(E) \\
\bm{B}(E)
\end{array} \right]   \label{BEM-IE1}
\eeq
and $\bm{u}_{\I}$ represents the vector encompassing 2$N_n$ unknown nodal
degrees of freedom $\psi^{\I}$ and $\psi^{\I}_{,n}$ on the interface $\Ga$.

Non-trivial solutions for $\bm{u}_{\I}$ in Eq. (\ref{BEM-IE}) can be
determined from the following condition:
\beq
\mbox{det}\Big[\bm{C}(E)\Big] = 0   \label{BEM-IE2}
\eeq

The roots $E$ of the above equation represent the energy eigenvalues of the 
quantum system. By using these roots in Eq. (\ref{BEM-IE}), non-trivial 
$\bm{u}_{\I}$ (eigenvectors) can be found (see, \eg, \cite{RM-book}). A 
specific sub-vector of $\bm{u}_{\I}$ containing exclusively the nodal 
wavefunctions $\psi$ can be utilized to compute $|\psi(Q)|^2$, which 
conveys the probability density associated with locating an electron at a 
precise spatial coordinate denoted as $Q$. It is important to note that the 
probability density should be appropriately normalized to ensure that the
total probability of finding the electron in the specified region is
100\%. In other words,
\beq
\int_V |\psi(Q)|^2\,\D Q = 1
\eeq
where $V$ includes both the structure and the region of interest within the 
barrier.

\subsection*{2.3. Conventional energy eigenvalue search}

When $\bm{C}(E)$ undergoes numerical evaluation, its determinant does not 
precisely equal zero for the roots $E$ (see Eqs. (\ref{BEM-IE1}) and 
(\ref{BEM-IE2})). Consequently, the common approach for 
identifying the real roots of Eq. (\ref{BEM-IE1}) involves searching for the 
local minima of $|\det[\bm{C}(E)]|$ within intervals where 
$E_{\m} \le E \le E_{\M}$, employing a small step size of $\Del E$. In this 
scenario, the number of search iterations can be expressed as 
$N_E=(E_{\M}-E_{\m})/\Del E+1$. Since $\bm{C}(E)$ is dependent on $E$, it is 
customary to opt for a very small value of $\Del E$, resulting in a notably 
high value of $N_E$, in order to obtain highly accurate results and prevent 
any eigenvalues from being 
overlooked during the scanning process. As a result, this computational 
technique (conventional method) is widely recognized for its significant 
computational cost, as it necessitates the re-evaluation of the boundary 
integrals specified in Eqs. (\ref{BIE-I}) and (\ref{BIE-E}), along with the 
determinant in Eq. (\ref{BEM-IE1}), at each iteration, amounting to a total
of $N_E$ iterations.

\subsection*{2.4. Accelerated energy eigenvalue search}

In this research, a novel approach to expedite the search for energy 
eigenvalues was introduced, inspired by techniques previously
developed in \cite{Phan-21} and \cite{Kari-21}.

For the interior region, as the arguments $k_{\I}r$ of $Y_0$ and $J_0$ are
real numbers (see Eq. (\ref{GI})), the technique presented in \cite{Phan-21}
and \cite{Kari-21} can be applied to rewrite the BIE (\ref{BIE-I}) as follows:
\beq
c(P)\psi^{\I}(P) = \sum_{j=0}^{\inf} \la_{\I}^j\int_{\Ga}\Big[
\bar{G}_{\I j}(P,Q)\,\psi^{\I}_{,n}(Q) - \bar{G}_{\I j,n}(P,Q)\,
\psi^{\I}(Q)\Big]\,\D Q      \label{BIE-I1}
\eeq
where $\la_{\I}=k_{\I}^2$, and $\bar{G}_{\I j}$ and $\bar{G}_{\I j,n}$
are defined in Eqs. (\ref{Gb1}) and (\ref{Gb1_n}), respectively.

The advantage of Eq. (\ref{BIE-I1}) over Eq. (\ref{BIE-I}) lies in the fact 
that the boundary integrals in Eq. (\ref{BIE-I1}) remain independent of the 
energy eigenvalues $E$ that are being sought.

Nonetheless, the aforementioned technique cannot be applied to the BIE 
(\ref{BIE-E}). This is due to the fact that, for the exterior region, the 
arguments $ik_{\E}r$ of $Y_0$ and $J_0$ are pure imaginary numbers (as
indicated in Eq. (\ref{GE})), invariably resulting in $J_0 > 0$. Consequently, 
this condition contradicts the prerequisite of $J_0 = 0$ as mandated for the 
applicability of the technique introduced in \cite{Phan-21} and 
\cite{Kari-21} in the context of the exterior region.

In the following, a novel technique is introduced to render BIE (\ref{BIE-E})
independent of $E$. By incorporating Eq. (\ref{GE}) into Eq. (\ref{BIE-E}), 
one gets
\beq
c(P)\psi^{\E}(P) =-{1\over 4}\int_{\Ga} \Big[(Y_0(ik_{\E}r)-iJ_0(ik_{\E}r))\,
\psi^{\E}_{,n}(Q) - (Y_{0,n}(ik_{\E}r)-iJ_{0,n}(ik_{\E}r))\,\psi^{\E}(Q)
\Big]\,\D Q      \label{BIE-E1}
\eeq

Consider the series expansions of $Y_0$ and $J_0$ given by
\beq
Y_0(ik_{\E}r) = {2\over\pi}\Bigg[\sum_{j=0}^{\inf} {\la_{\E}^j\,r^{2j}\over 
(-4)^j(j!)^2}\Big( \ln r - S_j + \ga+\ln{ik_{\E}\over 2} \Big) \Bigg]
   \label{Y0}
\eeq
and
\beq
J_0(ik_{\E}r) = \sum_{j=0}^{\inf}{\la_{\E}^j\,r^{2j}\over (-4)^j(j!)^2}   
   \label{J0}
\eeq
where $\la_E=(ik_{\E})^2=-k_{\E}^2$ and $\ga$ denotes the Euler-Mascheroni 
constant.

By letting
\beq
Y_0(ik_{\E}r) = Y_{01}(ik_{\E}r) + Y_{02}(ik_{\E}r)   \label{Y0_12}
\eeq
where
\beq
Y_{01}(ik_{\E}r) = {2\over\pi}\Bigg[\sum_{j=0}^{\inf}{\la_{\E}^j\,r^{2j}\over 
(-4)^j(j!)^2}\Big( \ln r - S_j \Big) \Bigg]   \label{Y_01}
\eeq
and
\beq
Y_{02}(ik_{\E}r) = {2\over\pi}\Big( \ga+\ln{ik_{\E}\over 2} \Big)
\sum_{j=0}^{\inf} {\la_{\E}^j\,r^{2j}\over (-4)^j(j!)^2}   \label{Y_02}
\eeq
Eq. (\ref{GE}) becomes
\beq
G_{\E}(P,Q) =-{1\over 4}\Big(Y_{01}(ik_{\E}r) - iJ_0(ik_{\E}r) + 
Y_{02}(ik_{\E}r) \Big) = G_{\E 1}(P,Q) + G_{\E 2}(P,Q)   \label{GE0}
\eeq
where
\beq
G_{\E 1}(P,Q) =-{1\over 4}\Big(Y_{01}(ik_{\E}r) - iJ_0(ik_{\E}r) \Big) =
\sum_{j=0}^{\inf} \la_{\E}^j\;{r^{2j}\Big[2(S_j - \ln r) + i\pi\Big]\over 
4\pi(-4)^j(j!)^2}   \label{GE1}
\eeq
and
\beq
G_{\E 2}(P,Q) =-{1\over 4}Y_{02}(ik_{\E}r) =
\Big( \ga+\ln{ik_{\E}\over 2} \Big)\sum_{j=0}^{\inf} \la_{\E}^j\;{-r^{2j}\over
2\pi(-4)^j(j!)^2}   \label{GE2}
\eeq

By employing Eqs. (\ref{Y0_12}) through (\ref{GE2}), Eq. (\ref{BIE-E1}) is
transformed into
\bea
c(P)\psi^{\E}(P) &=& \sum_{j=0}^{\inf} \la_{\E}^j\Bigg\{ 
\int_{\Ga}\Big[\bar{G}_{\E 1j}(P,Q)\,\psi^{\E}_{,n}(Q) - 
\bar{G}_{\E 1j,n}(P,Q)\,\psi^{\E}(Q)\Big]\,\D Q \nonumber \\[1mm]
&+& \eta\int_{\Ga}\Big[\bar{G}_{\E 2j}(P,Q)\,\psi^{\E}_{,n}(Q) - 
\bar{G}_{\E 2j,n}(P,Q)\,\psi^{\E}(Q)\Big] \,\D Q \Bigg\}    \label{BIE-11}
\eea
where
\bea
\eta &=& \ga + \ln {ik_{\E}\over 2}   \\[2mm]
\bar{G}_{\E 1j}(P,Q) &=& \bar{G}_{\I j}(P,Q) = 
{r^{2j}\Big[2(S_j - \ln r) + i\pi\Big]\over 4\pi(-4)^j(j!)^2}
   \label{Gb1} \\[2mm]
\bar{G}_{\E 2j}(P,Q) &=& {-r^{2j}\over 2\pi(-4)^j(j!)^2} 
\label{Gb2} \\[2mm]
\bar{G}_{\E 1j,n}(P,Q) &=& \bar{G}_{\I j,n}(P,Q) = 
{r^{2j-1}\Big[2j(S_j - \ln r) - 1 + ij\pi\Big]
\over 2\pi(-4)^j(j!)^2}\,{\pa r\over\pa\bm{n}}   \label{Gb1_n}  \\[2mm]
\bar{G}_{\E 2j,n}(P,Q) &=& {-jr^{2j-1}
\over \pi(-4)^j(j!)^2}\,{\pa r\over\pa\bm{n}}   \label{Gb2_n} 
\eea

In contrast to Eqs. (\ref{GI}) and (\ref{GE}), the fundamental solutions 
described above are shown to be independent of the wavenumbers $k_{\I}$ 
and $k_{\E}$, and thereby, independent of the energy $E$.

By discretizing Eqs. (\ref{BIE-I1}) and (\ref{BIE-11}) with $N$ boundary 
elements having a total of $N_n$ nodes, and using the 
first $(m+1)$ terms of the series expansion on the right hand side of
these equations, the following equations are respectively obtained for
each source point $P=1 \ldots N_n$:
\bea
\sum_{j=0}^{m} \la_{\I}^j \Bigg( \sum_{k=1}^{N_n} \Gb^{\I}_{jPk}
\psi^{\I}_{,nk} - \sum_{k=1}^{N_n} \Hb^{\I}_{jPk}\psi^{\I}_k \Bigg)
&=& 0   \label{BEA-I} \\
\sum_{j=0}^{m} \la_{\E}^j \Bigg( \mu\sum_{k=1}^{N_n} \Gb^{\E 1}_{jPk}
\psi^{\I}_{,nk} - \sum_{k=1}^{N_n} \Hb^{\E 1}_{jPk}\psi^{\I}_k +
\eta\Big( \mu\sum_{k=1}^{N_n} \Gb^{\E 2}_{jPk}\psi^{\I}_{,nk} -
\sum_{k=1}^{N_n} \Hb^{\E 2}_{jPk}\psi^{\I}_k\Big) \Bigg) &=& 0
   \label{BEA-E}
\eea
where Eqs. (\ref{BC1}) and (\ref{BC2}) are used in Eq. (\ref{BEA-E})
to convert $\psi^{\E}_{,nk}$ and $\psi^{\E}_k$ into $\psi^{\I}_{,nk}$ and
$\psi^{\I}_k$, respectively.

When applied to all $N_n$ source points, the two equations above form two 
linear systems, each containing $N_n$ equations. These systems can be 
represented in matrix form as follows:
\bea
\bm{\Ab}(E)\bm{u}_{\I} &=& \bm{0}   \label{BIE-num-a1}   \\[2mm]
\bm{\Bb}(E)\bm{u}_{\I} &=& \bm{0}   \label{BIE-num-b1}
\eea
where
\bea
\bm{\Ab}(E) &=& \sum_{j=0}^{m} \la_{\I}^j\,\bm{\Ab}_j   \label{BIE-num-a2}
   \\[1mm]
\bm{\Bb}(E) &=& \sum_{j=0}^{m} \la_{\E}^j\,\Big(\bm{\Bb}_{1j} + 
\eta\bm{\Bb}_{2j}\Big)   \label{BIE-num-b2}
\eea

The condition for having non-trivial solutions for $\bm{u}_{\I}$ is
\beq
\mbox{det}\Big[\bm{\Cb}(E)\Big] = 0   \label{det_C_bar}
\eeq
where
\beq
\bm{\Cb}(E) = \left[ \begin{array}{c}
\bm{\Ab}(E) \\
\bm{\Bb}(E)
\end{array} \right]   \label{C_bar}
\eeq

As indicated by Eqs. (\ref{BIE-num-a2}) and (\ref{BIE-num-b2}), 
$\bm{\Ab}$ and $\bm{\Bb}$ are polynomials whose
coefficients $\bm{\Ab}_j$ (for $\bm{\Ab}$) and $\bm{\Bb}_{1j}$ and
$\bm{\Bb}_{2j}$ (for $\bm{\Bb}$) are independent of $E$ (the integrals 
in Eqs. (\ref{BIE-I1}) and (\ref{BIE-11}) are not functions of $E$). This 
means that, for a given
problem, these boundary integrals only need to be evaluated $(m+1)$ times
to determine $3(m+1)$ coefficients $\bm{\Ab}_0, \bm{\Ab}_1, \ldots, 
\bm{\Ab}_m$, $\bm{\Bb}_{10}, \bm{\Bb}_{11}, \ldots, \bm{\Bb}_{1m}$, and
$\bm{\Bb}_{20}, \bm{\Bb}_{21}, \ldots, \bm{\Bb}_{2m}$.
Then, as polynomials (see Eqs. (\ref{BIE-num-a2}) and (\ref{BIE-num-b2})), 
$\bm{\Ab}$ and $\bm{\Bb}$, thereby $\bm{\Cb}$
can be rapidly computed for every value of $E$ over the range
$0 < E < V_0$, even with a very small increment $\Del E$. It should be noted 
that $m$ is typically much smaller that than $N_E$. Consequently, the 
proposed technique enhances the computational efficiency of searching for 
the local minima of $|\mbox{det}[\bm{\Cb}(E)]|$ compared to the
conventional approach using $|\mbox{det}[\bm{C}(E)]|$ as described in 
Section 2.3.

\section*{3. Numerical Examples}

The accuracy and efficiency of the proposed approach were evaluated in this 
study by analyzing four case studies: the first case has an analytical 
solution, while the last three had previously been explored by 
other research teams. These cases centered on straight quantum wires 
characterized by uniform cross-sections. MATLAB scripts were devised to 
numerically implement the proposed method employing quadratic boundary 
elements. In the case of the first three examples, a value of $m=30$ is 
adequate for accurate results. The advantage of the proposed technique is
represented by its gain, defined as the ratio of the computational
times required by the conventional and the proposed methods.

\subsection*{3.1. A quantum wire of circular cross-section}

Consider the problem of a GaAs quantum wire embedded in an infinite 
Ga$_{0.63}$Al$_{0.37}$As barrier region, which was previously studied in 
\cite{Gospavic-08}. This wire features a circular cross-section with a radius 
of 5 nm. The parameters for this quantum structure are as follows: 
$m_{\I}^*=0.0665m_0$, $m_{\E}^*=0.0858m_0$, $V_0=320$ meV where $m_0$ denotes
the free electron mass.

In this study, we discretized the circular boundary using 24 uniform 
quadratic boundary elements. To achieve numerical precision to four decimal 
places, we selected a value of $\Delta E=10^{-4}$ meV. For the energy interval 
of $1 \le E \le 319$ meV, this necessitated repeating the BEA (conventional 
method) a total of 3,180,001 times, making it a very time-consuming process.

\begin{figure}[!h]
\centerline{\includegraphics[height=10cm]{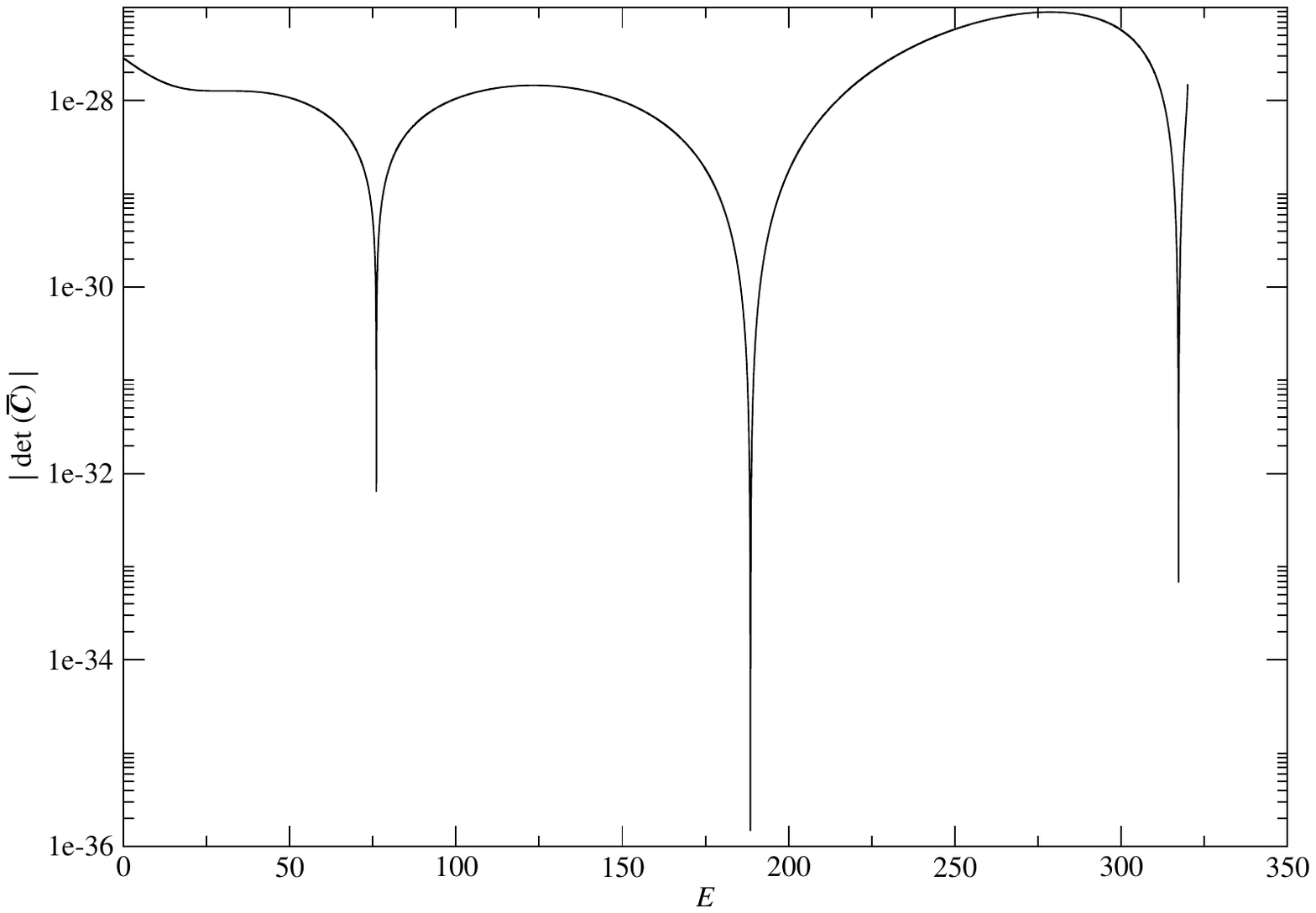}}
\caption{\label{Circ_iter} Plot of $|\mbox{det}[\bm{\bar{C}}(E)]|$ for the
quantum wire of circular cross-setion.}
\end{figure}

\begin{table}[!h]
\centering
\begin{tabular}{ccc}
\hline \hline
&& \\[-2mm]
Analytical                     &    Proposed      &  Conventional     \\
 Solution \cite{Harrison-book} &     Method       &     Method        \\
\hline
&& \\[-2mm]
 76.0663  & 76.0610 (0.007\%) & 76.0605 (0.008\%) \\
188.5605  &188.5511 (0.005\%) &188.5512 (0.005\%) \\
317.2923  &317.2861 (0.002\%) &317.2878 (0.001\%) \\
\hline 
\end{tabular}
\caption{\label{table1} The energy eigenvalues $E$ (in meV) for the quantum wire
of circular cross-section. The percentage errors are given in parentheses.}
\end{table}

Figure \ref{Circ_iter} illustrates the local minima of 
$|\mbox{det}[\bm{\bar{C}}(E)]|$ within the specified interval. All three 
energy eigenvalues, as determined by the proposed method, demonstrate 
excellent agreement with both the analytical results \cite{Harrison-book} and 
those obtained using the conventional method (see Table \ref{table1}). Notably, 
it is important to highlight that the gain in this case was approximately 1,316.

In the case of a quantum structure exhibiting specific symmetries, such as the 
quantum wire discussed in this section, the presence of degenerate energy levels 
may arise. When employing either conventional or proposed methods, it is
necessary to conduct further analysis on the wavefunctions corresponding to a 
given energy eigenvalue. If multiple linearly independent wavefunctions are 
associated with the same energy eigenvalue, the quantum state at that energy 
level is considered degenerate. For this problem, the 
analytical solution \cite{Harrison-book} indicates that the two
quantum states at 188.5605 meV and 317.2923 meV are doubly degenerate.

\begin{figure}[!htp]
  \hspace{-1cm}
  \subfigure[]{\label{Circ_Dens_2}\includegraphics[width=9cm]{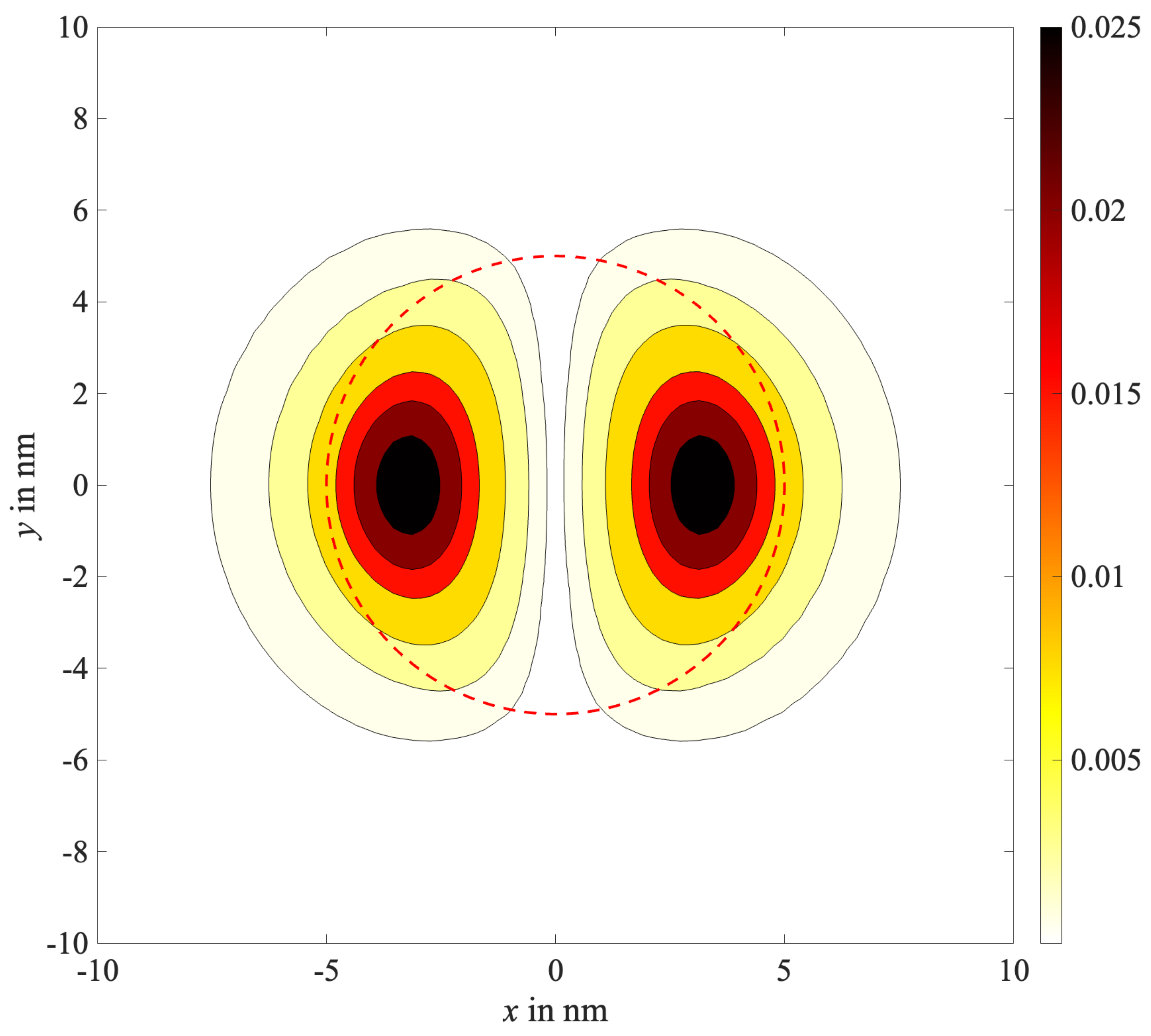}} 
  \subfigure[]{\label{Circ_Dens_3}\includegraphics[width=9cm]{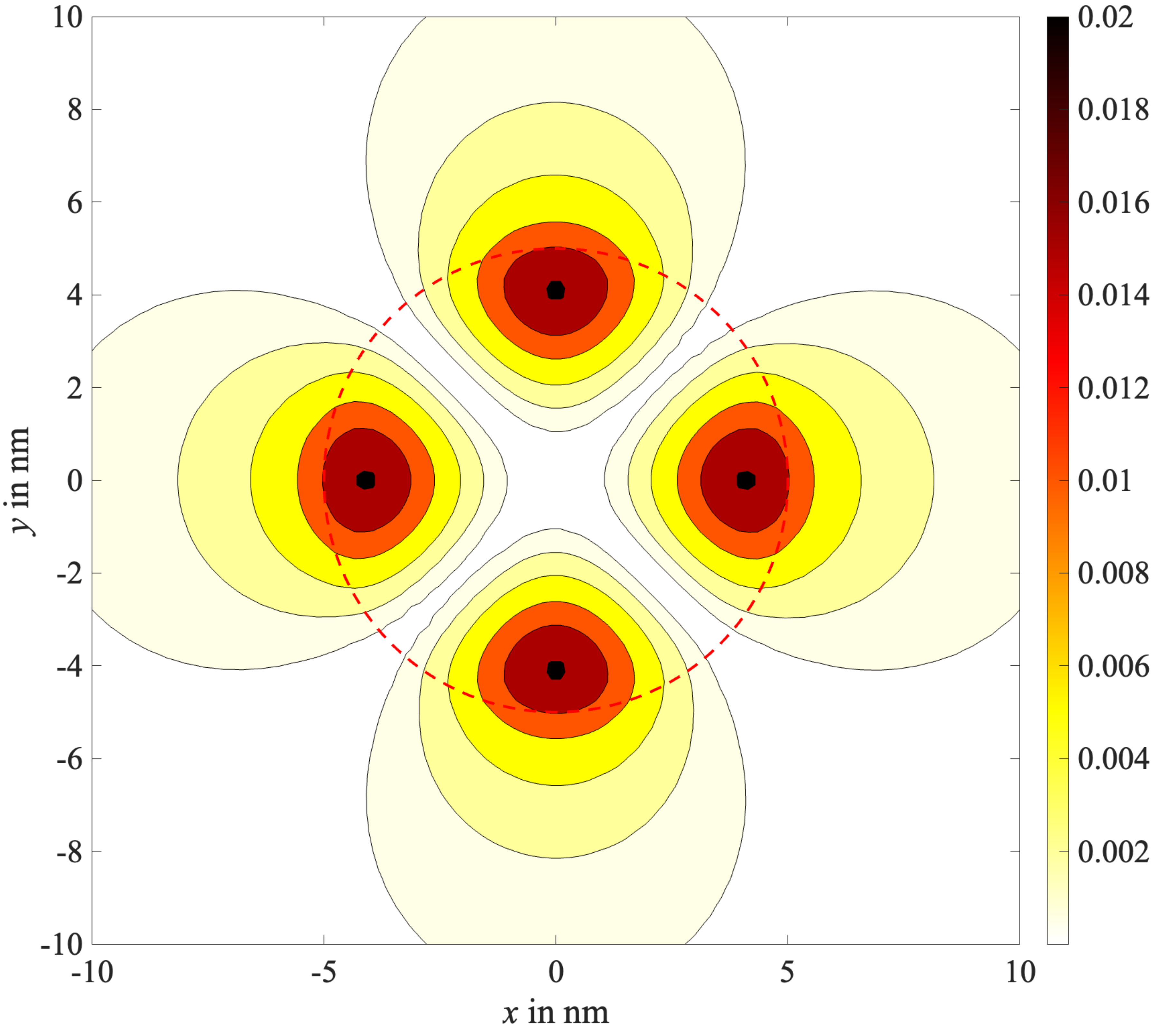}} \\[-4mm]
\caption{Contour plots of the normalized probability density (in nm$^{-2}$) for 
the two highest excited states in a quantum wire of circular cross-section 
(dashed line) at energy levels of (a) $E=188.551$ meV, and (b) $E=317.286$ meV.}
\end{figure}

The next phase in evaluating the proposed technique involves the examination 
of contour plots depicting the normalized probability density. These plots 
offer valuable insights into the spatial distribution of electrons at
distinct energy levels. Figures
\ref{Circ_Dens_2} and \ref{Circ_Dens_3} display contour plots of the
normalized probability density for the two highest excited states.
In subplot \ref{Circ_Dens_2}, at 
an energy level of 188.551 meV, we observe a distinctive pattern, 
indicating the preferential localization of electrons within the quantum wire.
This pattern changes noticeably in subplot \ref{Circ_Dens_3}, corresponding 
to an energy level of 317.286 meV, reflecting the influence of higher energy 
states. The contour plots provide a comprehensive view of how the 
wavefunctions evolve with energy, offering a deeper understanding of the 
quantum behavior within the wire.

It's important to note that these 
contour plots can have an infinite number of orientations due to the 
continuous rotational symmetry inherent in the circular cross-section. Each 
orientation corresponds to a distinct eigenvector of the nodal normalized 
wavefunctions. The orientation shown in Figs. \ref{Circ_Dens_2} and 
\ref{Circ_Dens_3} corresponds to an eigenvector in which the first nodal
wavefunction was normalized to a unit value. This particular type of 
eigenvector also applies to the subsequent contour plots presented in this 
work.

\subsection*{3.2. A quantum wire of square cross-section}

This example, as explored in reference \cite{Gelbard-01}, features a wire 
with a square cross-section, each side measuring 10 nm. The
effective masses for both the structure and barrier are $m_{\I}^*=m_{\E}^*=
0.06m_0$, and the potential barrier is $V_0=300$ meV. A total of 20 uniform
quadratic boundary elements was employed in this work to mesh the square 
boundary, and the numerical results with three decimal digits were achieved 
using $\Del E=10^{-3}$ meV. By scanning $|\det[\bm{\bar{C}}(E)]|$ within the
$1 \le E \le 299$ meV using this step size $\Del E$, three local minima of
$|\det[\bm{\bar{C}}(E)]|$ at energy eigenvalues 73.897 meV, 178.190 meV
and 274.008 meV were identified as depicted in Fig. \ref{Circ_iter}. 

\begin{figure}[!h]
\centerline{\includegraphics[height=10cm]{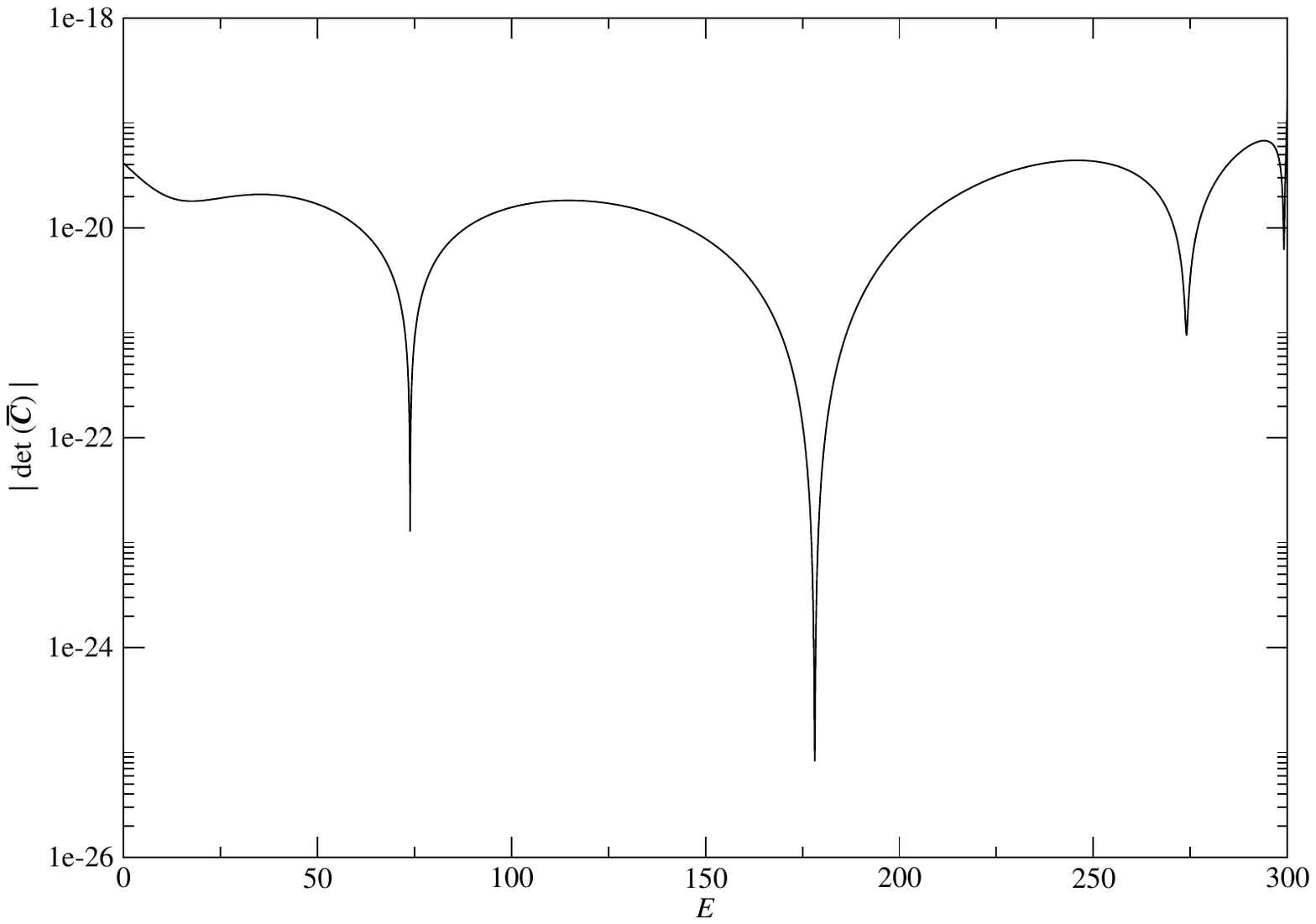}}
\caption{\label{Square_iter} Plot of $|\mbox{det}[\bm{\bar{C}}(E)]|$ for the
quantum wire of square cross-setion.}
\end{figure}

\begin{table}[!h]
\centering
\begin{tabular}{ccc}
\hline \hline
&& \\[-2mm]
Proposed   &Conventional     & Ref. \cite{Gelbard-01} \\
 Method    &   Method        &                        \\
\hline
&& \\[-2mm]
 73.897   & 73.885 (0.016\%) & 74.2 (0.410\%)    \\
178.190   &178.180 (0.006\%) &179   (0.455\%)    \\
274.008   &274.016 (0.003\%) &   -               \\
\hline 
\end{tabular}
\caption{\label{table2} The energy eigenvalues $E$ (in meV) for the quantum 
wire of square cross-section. The percentage discrepancies relative to the 
proposed method are indicated in parentheses.}
\end{table}

\begin{figure}[!htp]
  \hspace{-1cm}
  \subfigure[]{\label{Square_Dens_2}\includegraphics[width=9cm]{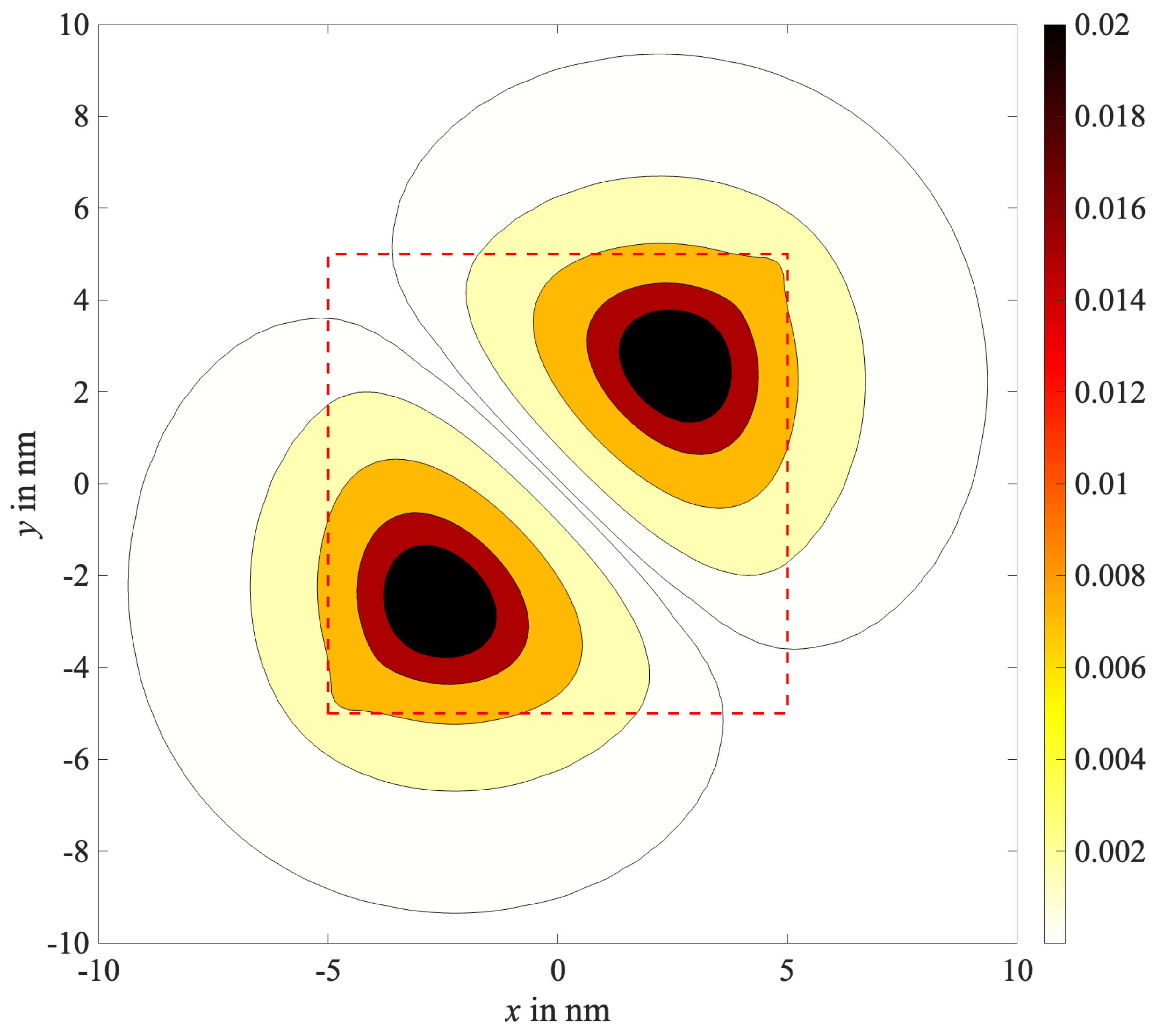}} 
  \subfigure[]{\label{Square_Dens_3}\includegraphics[width=9cm]{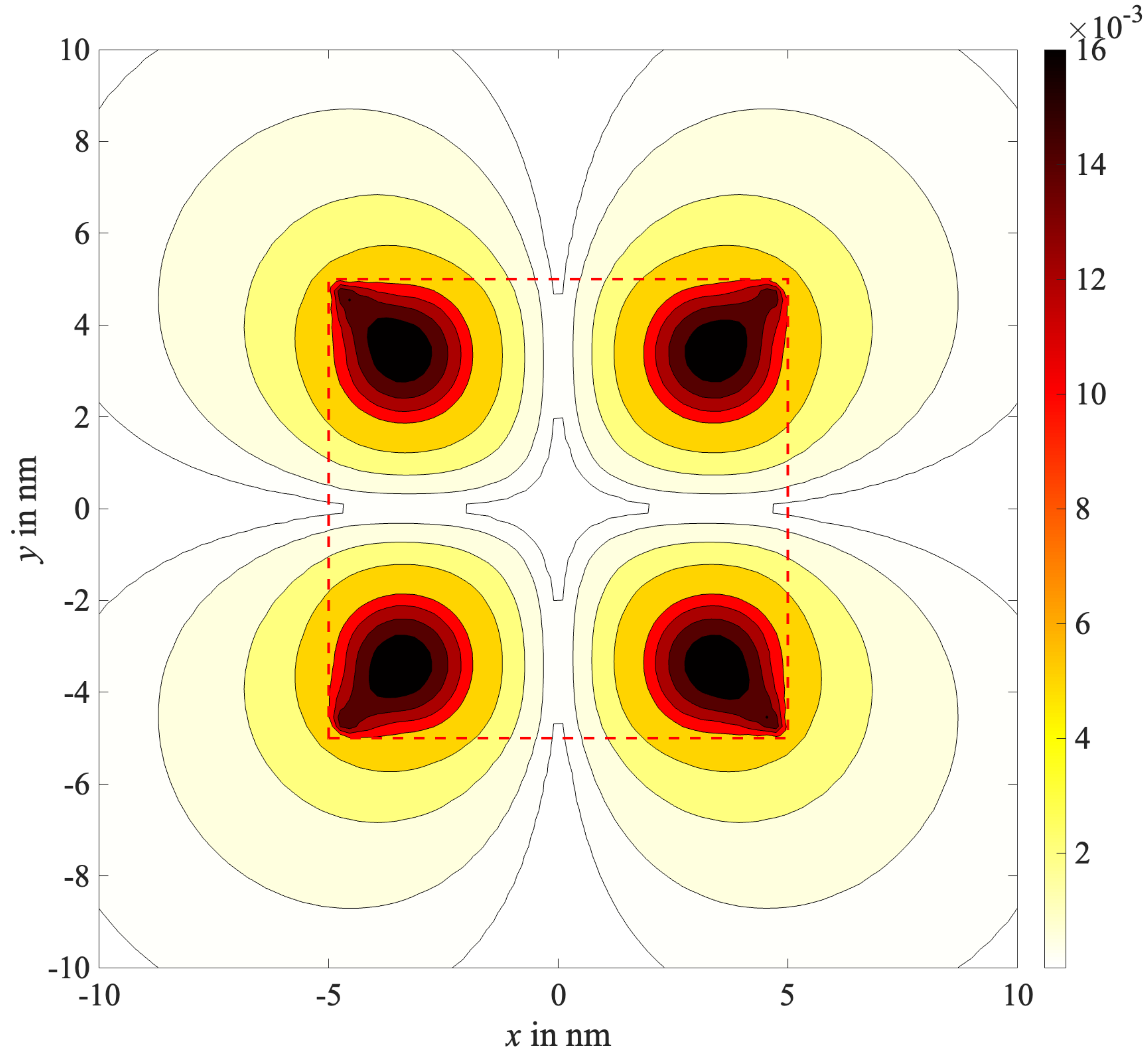}} \\[-4mm]
\caption{Contour plot of the normalized probability density (in nm$^{-2}$) for 
the two highest excited states in a quantum wire with a square cross-section 
(dashed line) at energy levels of (a) $E=178.190$ meV, and (b) $E=274.008$ meV.}
\end{figure}

As presented in Table \ref{table2}, the three energy eigenvalues obtained 
through the proposed method demonstrate a very high level of agreement with 
those derived from the conventional approach. While the highest energy 
eigenvalue was not provided in \cite{Gelbard-01}, numerical results from 
the proposed method are consistent with the two lowest energy eigenvalues 
reported in that reference. Furthermore, as mentioned in \cite{Gelbard-01},
178.190 meV is a repeated energy eigenvalue.
Notably, the proposed method exhibits remarkable computational efficiency, 
exemplified by the fact that, for this specific problem, the gain was
around 1,712.

The contour plots in Figs. \ref{Square_Dens_2} and \ref{Square_Dens_3}
illustrate the normalized probability density for the two highest excited 
states at energy levels of 178.190 meV and 274.008 meV within this quantum 
square wire. The contour plot in subplot \ref{Square_Dens_2} bears resemblance 
to the corresponding one found in \cite{Gelbard-01}, with the exception of 
the orientation of the lobes between the two plots. As noted in
\cite{Gelbard-01}, the excited state at 178.190 meV exhibits double degeneracy.
The vertically aligned lobes observed in the same reference were achieved by 
linearly combining two null vectors, each originating from one of the two 
degenerate states.

\subsection*{3.3. A quantum wire of rectangular cross-section}

In this section, the quantum wire case with a rectangular cross-section 
measuring 20 nm in width and 10 nm in height, as previously examined in 
\cite{Gospavic-08}, was revisited. The material properties remained 
consistent with those described in Section 3.1, and the barrier potential 
was maintained at 276 meV. For the BEA in this study, the rectangular 
interface was discretized using ten uniform elements along the longer side 
and five uniform elements along the shorter side, resulting in a total of 30 
quadratic boundary elements.

\begin{figure}[!h]
\centerline{\includegraphics[height=10cm]{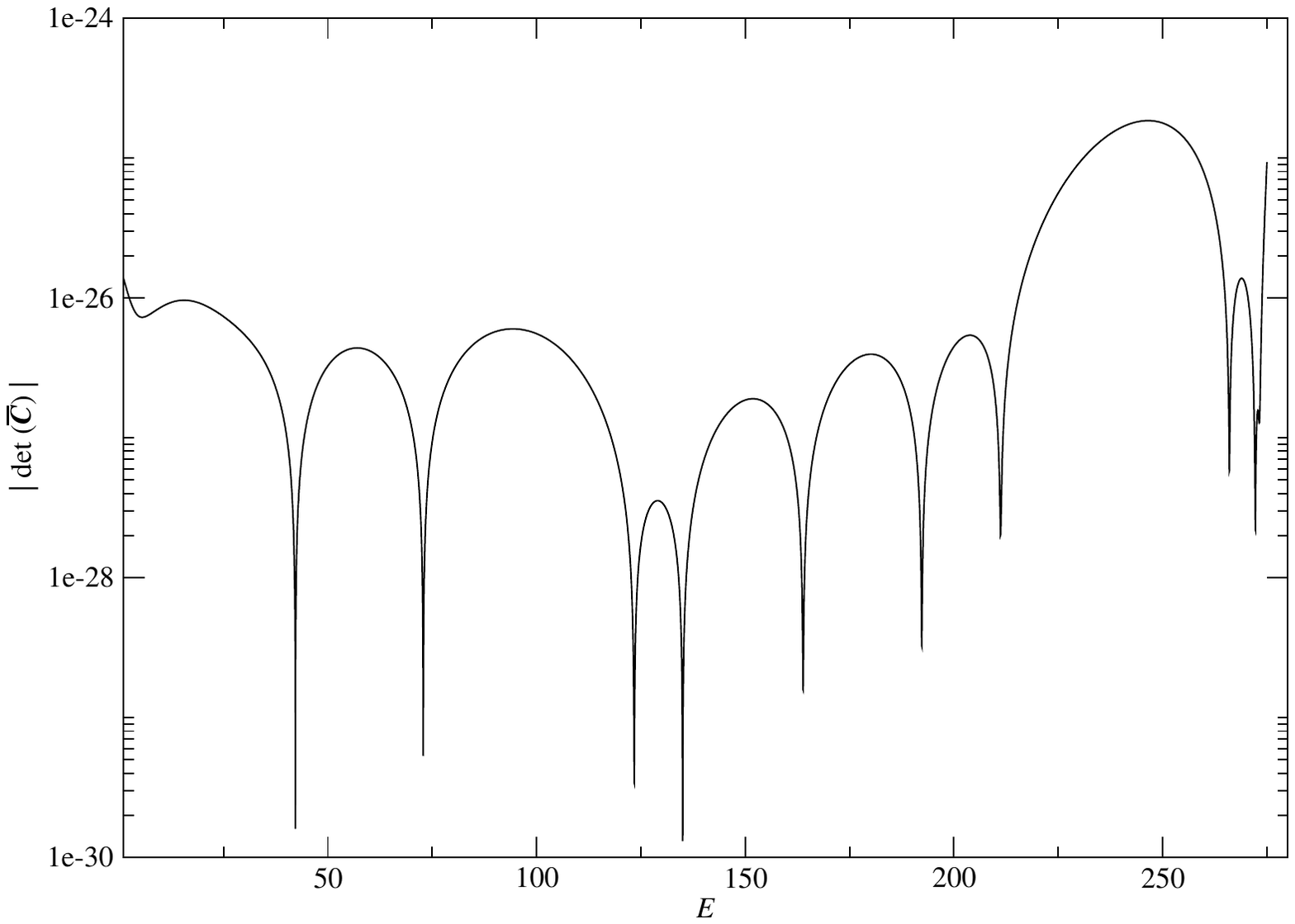}}
\caption{\label{Rect_iter} Plot of $|\mbox{det}[\bm{\bar{C}}(E)]|$ for the
quantum wire of rectangular cross-setion.}
\end{figure}

\begin{table}[!h]
\centering
\begin{tabular}{ccc}
\hline \hline
&& \\[-2mm]
Proposed & Ref.               & Ref.                \\
 Method  & \cite{Gospavic-08} & \cite{Pokatilov-00} \\
\hline
&& \\[-2mm]
 42.206  & 42.276 (0.166\%)   & 42.27  (0.151\%) \\
 72.849  & 73.101 (0.346\%)   &   -              \\
123.397  &123.899 (0.407\%)   &123.86  (0.375\%) \\
135.002  &135.177 (0.130\%)   &135.22  (0.162\%) \\
163.870  &164.636 (0.467\%)   &   -       \\
192.297  &193.065 (0.399\%)   &   -       \\
211.192  &212.717 (0.722\%)   &   -       \\
266.005  &266.552 (0.206\%)   &   -       \\
272.225  &272.801 (0.212\%)   &   -       \\
273.176  &275.049 (0.686\%)   &   -       \\
\hline 
\end{tabular}
\caption{\label{table3} The energy eigenvalues $E$ (in meV) for the 
quantum wire of rectangular cross-section. The percentage discrepancies 
relative to the proposed method are indicated in parentheses.}
\end{table}

Figure \ref{Rect_iter} depicts ten local minima of 
$|\mbox{det}[\bm{\bar{C}}(E)]|$ in the interval $1 \le E \le 275$ meV. 
The results displayed in Table \ref{table3} indicate a strong agreement 
between the numerical outcomes achieved through the proposed method and those 
documented in \cite{Gospavic-08} and \cite{Pokatilov-00}. These results 
exhibit a percentage discrepancy of less than 0.75\%. Moreover, the gain
for this particular case was roughly 1,828.

\begin{figure}[!h]
\centerline{\includegraphics[width=16cm]{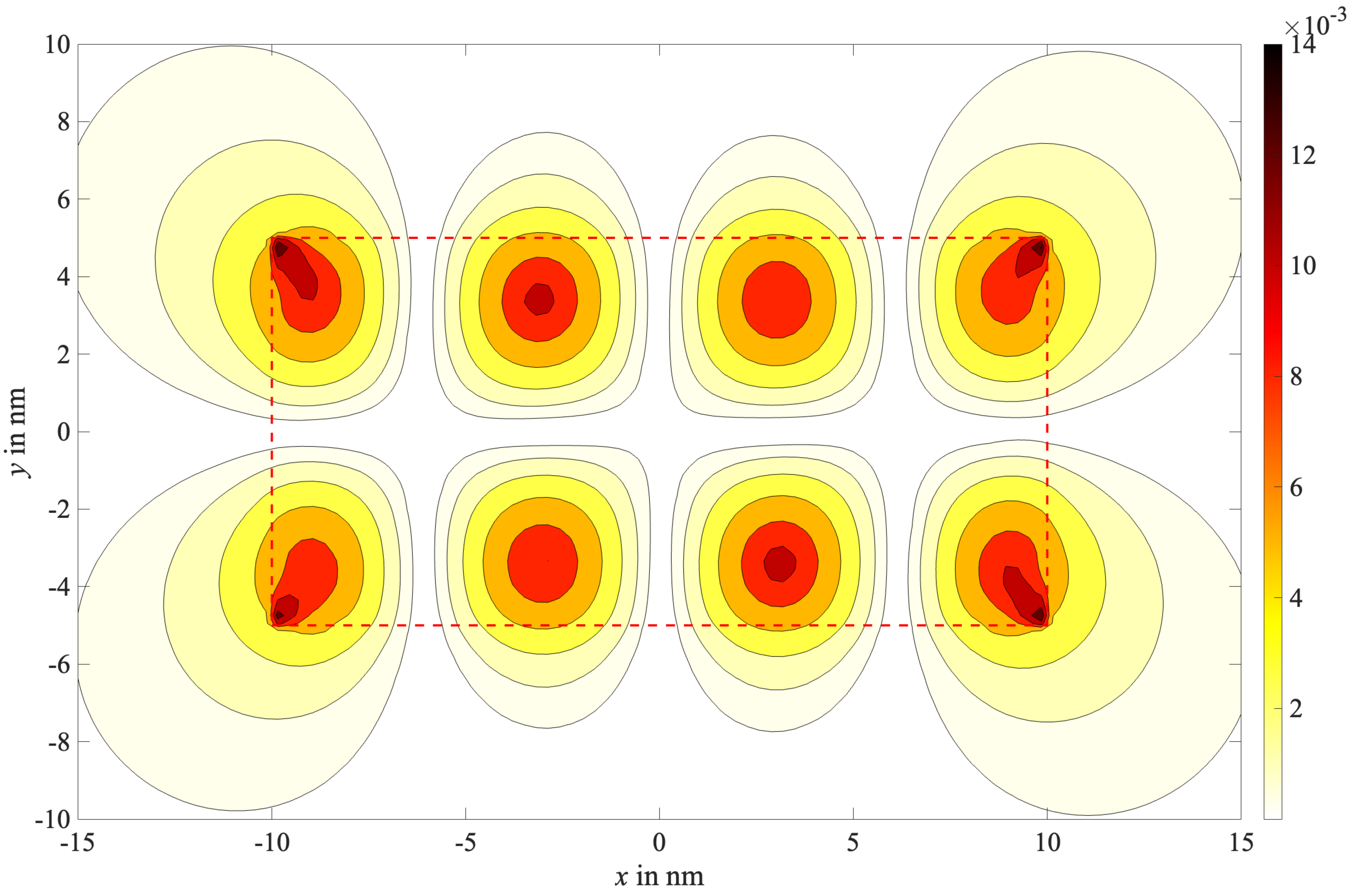}}
\caption{\label{Rect_Dens} Contour plot of the normalized probability density 
(in nm$^{-2}$) for the highest excited state in a quantum wire of rectangular
cross-section (dashed line) at energy level of $E=273.176$ meV.}
\end{figure}

In Fig. \ref{Rect_Dens}, a contour plot of the normalized probability density 
for the highest excited state at an energy eigenvalue of 273.176 meV is 
depicted, highlighting the presence of point symmetry about the origin
(0,0). 
 
\subsection*{3.4. A quantum wire of oval cross-section}

In this final example, a quantum wire with a stadium-shaped cross-section, a 
configuration previously examined in \cite{Knipp-96}, was reevaluated.
The stadium geometry presents a unique challenge for the analysis 
of energy eigenvalues, as it combines both curved and flat regions, creating 
a boundary that demands special consideration. In fact, $m=50$ had to be
selected to obtain converged numerical results. The structure and barrier 
both have effective masses of $0.0665m_0$, and the potential barrier has a 
height of $190$ meV.

For the given oval boundary, the radius of the semicircles is 12.5 nm, and 
the length of the line segments is 25 nm. Eight uniform quadratic elements 
discretize each of the semicircles, while each of the line segments employ 
four uniform quadratic elements.

\begin{figure}[!h]
\centerline{\includegraphics[height=10cm]{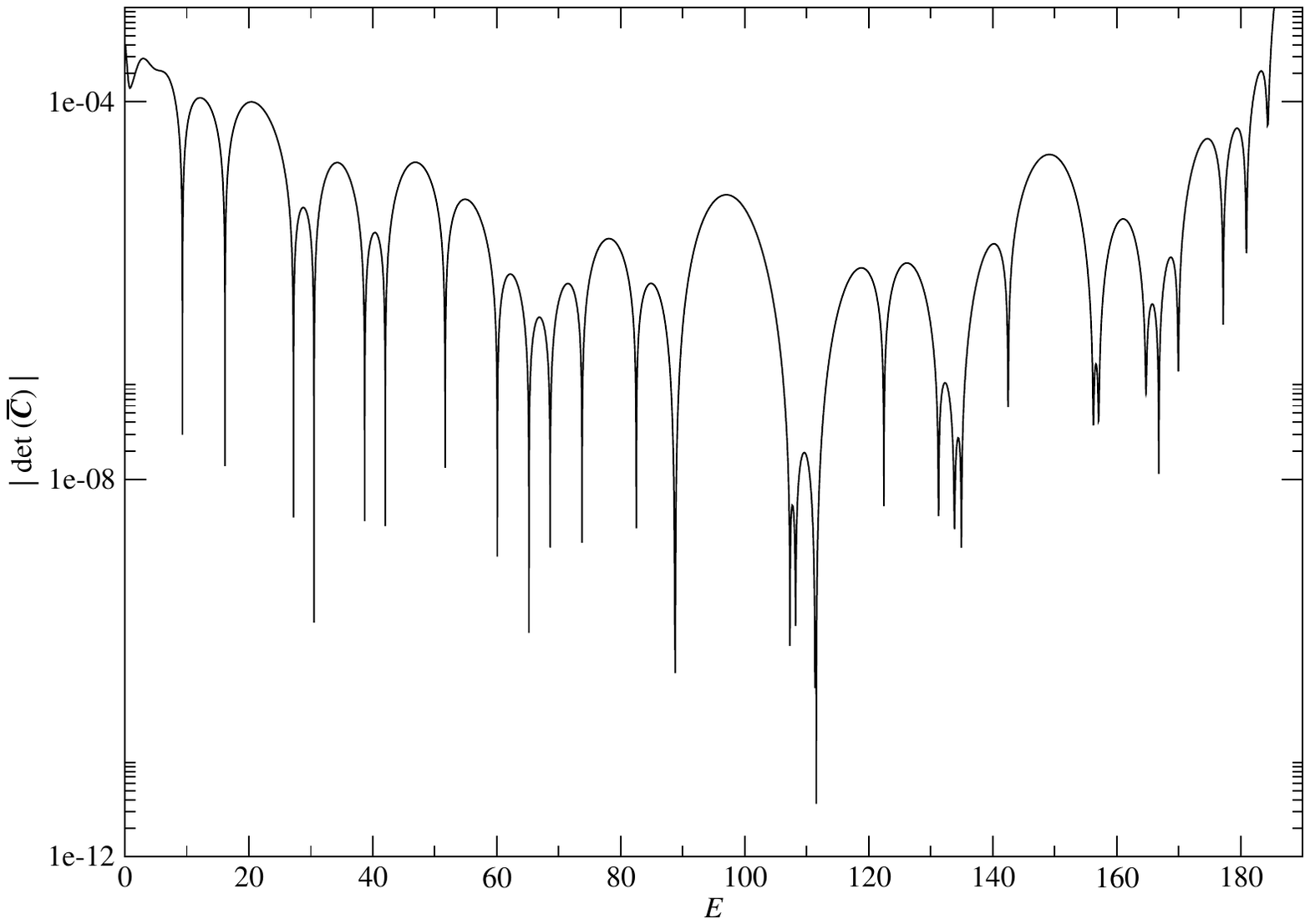}}
\caption{\label{Stadium_iter} Plot of $|\mbox{det}[\bm{\bar{C}}(E)]|$ for the
quantum wire of oval cross-setion.}
\end{figure}

\begin{table}[!h]
\centering
\begin{tabular}{rcc||rcc}
\hline \hline
&& \\[-2mm]
Proposed&  Conventional    &Ref. \cite{Knipp-96}&Proposed&  Conventional    &Ref. \cite{Knipp-96} \\
 Method &   Method         &                    & Method &     Method       &           \\
\hline
&& \\[-2mm]
  9.296 &  9.296 (0\%)     &         -          &111.539 &111.539 (0\%)     &   -       \\
 16.156 & 16.156 (0\%)     &         -          &122.442 &122.441 (0.001\%) &   -       \\
 27.197 & 27.197 (0\%)     &         -          &131.251 &131.249 (0.002\%) &   -       \\
 30.521 & 30.521 (0\%)     &         -          &133.823 &133.816 (0.005\%) &   -       \\
 38.700 & 38.700 (0\%)     &         -          &134.946 &134.938 (0.006\%) &   -       \\
 42.010 & 42.010 (0\%)     &         -          &142.486 &142.475 (0.008\%) &   -       \\
 51.665 & 51.663 (0.004\%) &         -          &156.239 &156.246 (0.004\%) &   -       \\
 60.087 & 60.085 (0.003\%) &         -          &157.082 &157.075 (0.004\%) &   -       \\
 65.188 & 65.188 (0\%)     &         -          &164.722 &164.704 (0.011\%) &   -       \\
 68.613 & 68.609 (0.006\%) &         -          &166.782 &166.781 (0.001\%) &   -       \\
 73.760 & 73.759 (0.001\%) &         -          &169.936 &169.935 (0.001\%) &   -       \\
 82.503 & 82.499 (0.005\%) &         -          &177.169 &177.173 (0.002\%) &   -       \\
 88.721 & 88.721 (0\%)     &         -          &180.919 &180.917 (0.001\%) &   -       \\
107.292 &107.287 (0.005\%) &         -          &184.373 &184.368 (0.003\%) &184.4 (0.015\%) \\
108.199 &108.194 (0.005\%) &         -          &                           &           \\
\hline 
\end{tabular}
\caption{\label{table4} The energy eigenvalues $E$ (in meV) for the quantum 
wire of oval cross-section. The percentage discrepancies 
relative to the proposed method are given in parentheses.}
\end{table}

A total of 31 energy eigenvalues are visible in Figure \ref{Stadium_iter} as 
the local minima of $|\det[\bm{\bar{C}}(E)]|$ within the interval $1 \le E 
\le 189$ meV. For this problem, the conventional method necessitated 985
times more computational time compared to the proposed approach (the gain
was approximately 985).

Table \ref{table4} reveals an excellent agreement between 
the numerical results obtained from the proposed and conventional methods, 
with relative discrepancies mostly below 0.005\%. The maximum relative 
discrepancy, occurring at the energy eigenvalue of 164.705 meV, is 0.011\%.
It is worth noting that only the highest energy eigenvalue was mentioned in 
\cite{Gelbard-01}, but this value is consistent with the one obtained from 
the proposed method, with a relative discrepancy of 0.015\%.

\begin{figure}[!h]
\centerline{\includegraphics[width=16cm]{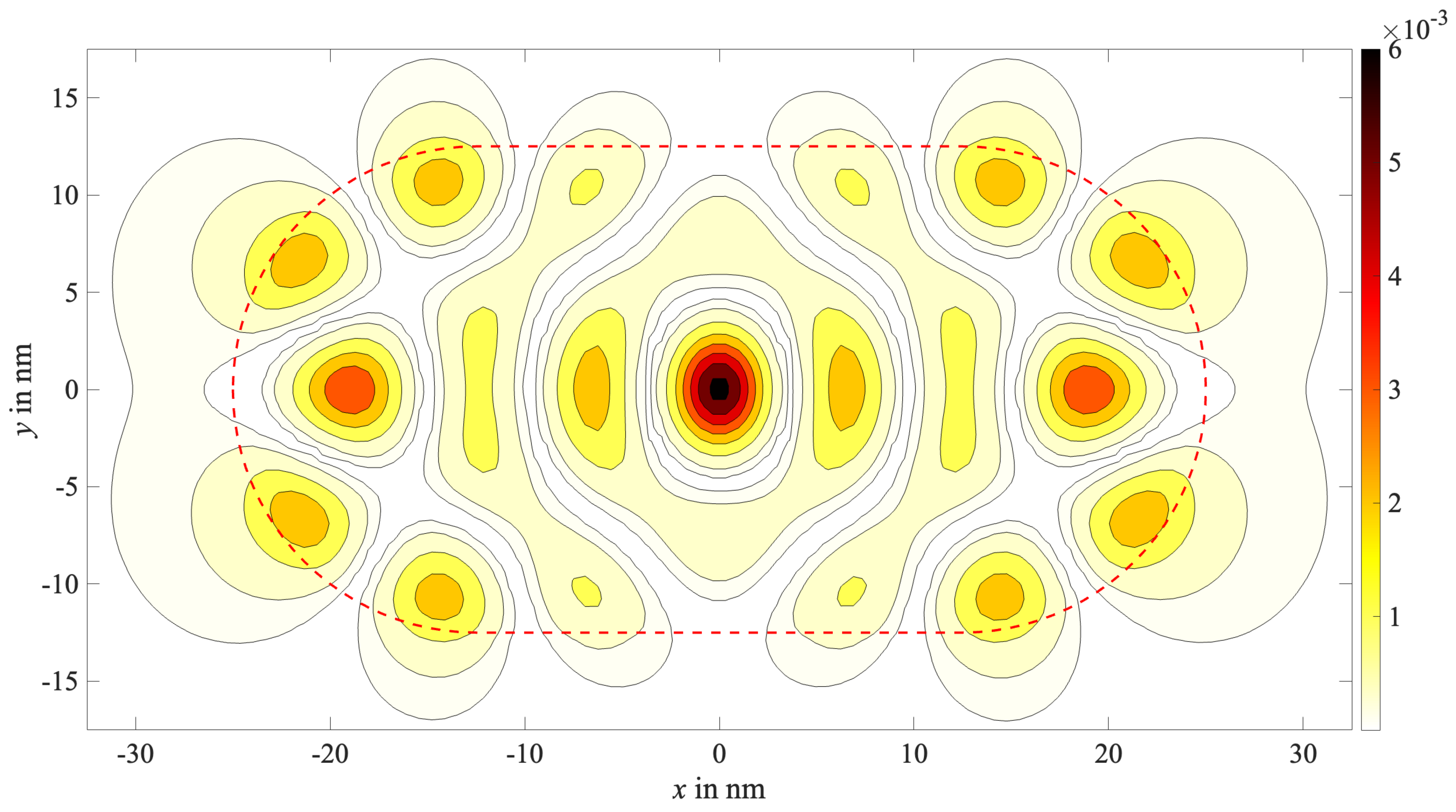}}
\caption{\label{Stadium_Dens} Contour plot of the normalized probability 
density (in nm$^{-2}$) for the highest excited state in a stadium quantum wire
(dashed line) at energy level of $E=184.373$ meV.}
\end{figure}

Figure \ref{Stadium_Dens} displays a contour plot depicting the normalized 
probability density for the highest excited state at the energy eigenvalue 
of 184.403 meV in the oval quantum wire. It is evident that 
this contour plot shares similarities with the one featured in \cite{Knipp-96}.

\section*{4. Conclusion}

This study introduces a highly efficient approach for computing energy 
eigenvalues in quantum semiconductor heterostructures. The accurate 
determination of electronic states within these heterostructures is of 
paramount importance for a comprehensive understanding of their optical and 
electronic properties, and as such, it represents a fundamental challenge in 
the field of semiconductor physics.

The novel approach presented in this work is rooted in the
utilization of series expansions of zero-order Bessel functions, which 
enables the numerical solution of the Schr\"{o}dinger equation using the 
boundary integral method for bound electron states in a manner that is not 
only computationally efficient but also highly accurate.

The credibility and utility of the proposed technique were rigorously tested 
on problems that have previously been investigated by other research groups.
The method's efficacy was demonstrated through the 
analysis of bound electron states in various quantum wire structures, 
highlighting its capability to handle complex potential profiles accurately 
and rapidly. The
presented fast boundary integral technique offers a promising tool for the 
design and optimization of semiconductor devices and paves the way for 
high-precision simulations of quantum electronic behavior in heterostructures.

In the case of multiply connected domains, Chen {\em et. al.} \cite{Chen-01} 
have demonstrated that even when utilizing a complex-valued fundamental 
solution, the BEA of the Helmholtz equation can produce spurious or fictitious 
eigenvalues. Notably, they found that these spurious eigenvalues are 
contingent upon the inner boundary of the domain and can be eliminated through 
the use of the Burton-Miller method \cite{BM-71}. However, to the best of the 
authors' knowledge, fictitious eigenvalues do not arise in multidomain models 
employed in the analysis of confined electron states within quantum structures.

\section*{Acknowledgements}

J. D. Phan acknowledges the Marilynn Sullivan Scholarship.


\end{document}